\documentclass[aps,prl,twocolumn,superscriptaddress,citeautoscript,floatfix,longbibliography]{revtex4-2}

\usepackage{physics}
\usepackage{graphicx}
\usepackage{gensymb}
\usepackage{bm, bbm, amssymb, siunitx}

\usepackage[normalem]{ulem}
\usepackage{textcomp}
\usepackage{xcolor}
\definecolor{mblue}{RGB}{0, 118, 186}

\definecolor{mgreen}{RGB}{29, 177, 0}

\definecolor{mred}{RGB}{209, 28, 36}

\usepackage{hyperref}
\hypersetup{
    colorlinks=true,
    linkcolor=blue,     
    urlcolor=blue,
    citecolor=blue
}
\setcitestyle{super,open={},close={}}

\newif\ifptitle
\newif\ifpnumber
\newcounter{para}
\newcommand\ptitle[1]{\par\refstepcounter{para}
{\ifpnumber{\noindent\textcolor{darkgray}{\textbf{\thepara}}\indent}\fi}
{\ifptitle{\textbf{[{#1}]}}\fi}}

\newif\iftrackchanges

\trackchangestrue  

\begin{document}

\title{Quantum-inspired design of a tunable broadband high-Q acoustic resonator}

\author{Jeffrey Shi}
\affiliation{Department of Physics, Harvard University, Cambridge, MA, 02138, USA}
\author{Benjamin H. November}
\affiliation{Department of Physics, Harvard University, Cambridge, MA, 02138, USA}
\author{Stephen Carr}
\affiliation{Brown Theoretical Physics Center and Department of Physics, Brown University, Providence, RI, 02912-1843, USA}
\author{Harris Pirie}
\affiliation{Department of Physics, Harvard University, Cambridge, MA, 02138, USA}
\affiliation{Clarendon Laboratory, University of Oxford, Oxford, OX1 3PU, UK}
\author{Jennifer E. Hoffman}
\affiliation{Department of Physics, Harvard University, Cambridge, MA, 02138, USA}
\affiliation{School of Engineering and Applied Sciences, Harvard University, Cambridge, MA, 02138, USA}

\date{\today}

\begin{abstract}
Resonators with a high quality factor (Q) are crucial components in a wide range of advanced technologies, including energy harvesting, chemical and biological sensing, and second-harmonic generation. Many applications also require resonance across a broad frequency range. However, single-cavity resonators face a fundamental trade-off between bandwidth and quality factor. Here we propose an acoustic resonator that overcomes this limitation by drawing inspiration from the ladder of harmonic oscillator states observed in twisted van der Waals heterostructures. By simulating an acoustic analog of twisted bilayer graphene, we discover a tunable ladder of acoustic resonances with Q as high as 4,000. These resonances are separated by as little as 10 Hz and persist over a bandwidth as broad as 1 kHz, forming an effective high-Q, broadband system. Our approach offers a promising pathway to overcome the inherent trade-offs in traditional resonators and paves the way for advanced high-Q acoustic devices.
\end{abstract}

\maketitle

\ptitle{Quantum-inspired design}
Acoustic metamaterials that adapt quantum concepts to the classical domain have led to applications such as topological insulator lasers \cite{harariS2018, bandresS2018}, acoustic logic gates \cite{XiaAM2018, PiriePRL2022}, and materials with zero or negative refractive index \cite{yangSA2021, duboisNC2017, tangNL2021}. Here, we design a broadband high-Q acoustic resonator by mimicking the dense array of flat bands found in a wide range of twisted semiconductor heterostructures\cite{wu2018, wu2019, xian2019}, including h-BN, MoSe$_2$, WSe$_2$, and WS$_2$~\cite{seyler2019, jin2019, peiPRX2022}. The relative twist between layers generates a long-range moir\'e pattern in these materials, resulting in a gradual variation of their interlayer alignment and the creation of a smoothly varying potential, $V(\mathbf{r})$. This potential changes the local bilayer Hamiltonian while leaving each monolayer band structure mostly intact~\cite{bistritzerPNAS2011, jung2014, carrNRM2020}. The flat bands of these twisted semiconductors thus have a straightforward explanation: the parabolic kinetic energy dispersion $E(\mathbf{k}) \propto |\mathbf{k}|^2$ of each monolayer band edge combines with the parabolic spatial potential of the moir\'e pattern $V(\mathbf{r}) \propto |\mathbf{r}|^2$ to give a textbook example of a harmonic oscillator~\cite{CarrPRR2020,angeliPNAS2021,tang2021}, characterized by the local Hamiltonian
\begin{align}
    H = \frac{|\mathbf{p}|^2}{2m^*} + \frac{1}{2}m^*\omega_0^2 |\mathbf{r}|^2,
    \label{eq:ho}
\end{align}
where $m^*$ is the effective electron mass and $\omega_0$ is the oscillator's frequency. This ladder of harmonic oscillator states differs from the well-known pair of flat bands in magic-angle twisted bilayer graphene\cite{caoN2018}, which arises from the hybridization of Dirac cones at fixed energy. 

\ptitle{Metamaterial Design} 
Here we design an acoustic metamaterial analog of twisted bilayer graphene where the harmonic oscillator flat bands manifest as multiple high-Q resonance modes with controllable frequency, number, and spacing. We start by designing two-dimensional acoustic metamaterials that mimic an electronic tight-binding model. Each lattice site consists of a cylindrical hole in a steel sheet, supporting single-cavity acoustic modes with frequency spacing set by the cavity size. We connect these modes \\
\begin{figure}[h!]
    \includegraphics[width=8.6cm]{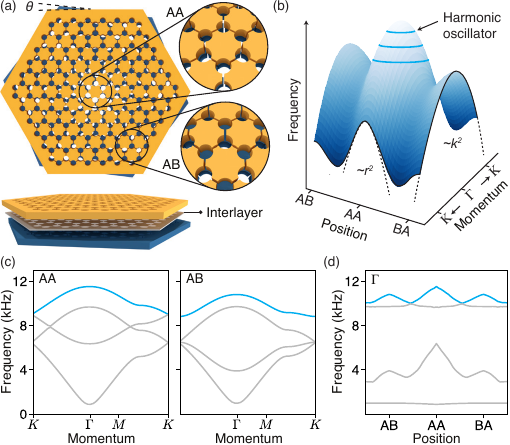}
    \caption{\textbf{A moir\'e potential generates an effective harmonic ocsillator Hamiltonian. a,} Acoustic metamaterial design to mimic twisted bilayer graphene. 
    \textbf{b,} Schematic of the band energy as a function of $\mathbf{k}$ and $\mathbf{r}$. At both AA and AB stacking the local band energy can be approximated by a parabola, leading to harmonic oscillator states near the local maxima.
    \textbf{c,} Calculated band structures for symmetric stacking configurations of acoustic bilayer graphene. At the $\Gamma$-point, the top band of the $s$-manifold (blue) is well approximated by a parabolic dispersion with negative curvature. \textbf{d,} $\Gamma$-point band edge frequency as a function of stacking configuration, calculated from an untwisted bilayer metamaterial with a lateral translation between the layers. This real-space band dispersion constitutes an effective potential $V(\mathbf{r})$ for the acoustic waves, which is locally parabolic at AA and AB. 
    }
\label{fig:intro}
\end{figure}
\begin{figure*}[t!]
\includegraphics[width=\textwidth]{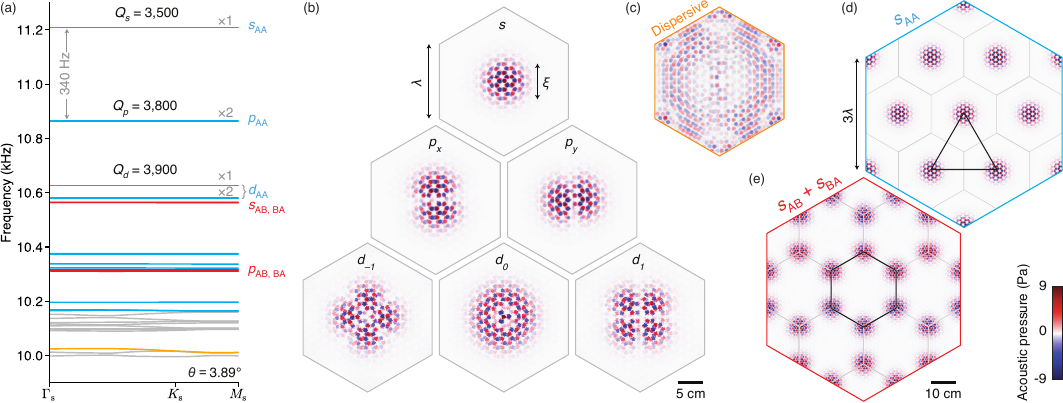}
\caption{\textbf{Harmonic oscillator potential generates a ladder of high-Q resonances. a,} Acoustic band structure of a 3.89$\degree$ twisted metamaterial at frequencies around the $\Gamma$-point band edge at the top of the $s$-manifold. The AA-localized flat bands (blue) are approximately equally-spaced with $\omega_0 \approx$ 340 Hz. The flat bands are either one-fold ($\times$1) or two-fold ($\times$2) degenerate. Harmonic oscillator states also localize on the AB/BA symmetry regions (red). A representative dispersive band is highlighted in orange. \textbf{b,} Pressure distribution of the AA-centered harmonic oscillator states labeled according to their angular harmonics.
\textbf{c,} Real-space pressure distribution for the representative dispersive mode. \textbf{d-e,} Pressure distribution over several moir\'e periods, showing larger moir\'e crystals with an effective triangular (d) or honeycomb (e) lattice.}
\label{fig:aa_harmonic}
\end{figure*}

\noindent through thin channels to allow nearest-neighbor intralayer hopping $t_{||}$, and focus on the lowest ($s$-band) manifold of states, whose dispersion closely mimics that of monolayer graphene \cite{gardezi2M2021}. To implement an interlayer coupling strength $t_z$, we insert a thin steel membrane between two vertically stacked graphene metamaterial layers \cite{dorrellPRB2020}. The couplings $t_{||}$ and $t_z$ are controlled by the channel width or interlayer membrane thickness, respectively. For a resonator in the audible range, we choose a cavity radius of $R = 3.25$ mm, nearest-neighbor spacing $a = 10$ mm, channel width $w = 0.875$ mm, metamaterial layer thickness $D = 1$ mm, and interlayer thickness $T = 25$ $\mu$m.  Finally, we introduce a twist angle $\theta = 3.89\degree$ between the layers to generate an acoustic moir\'e superlattice analogous to twisted bilayer graphene, as shown in Fig.~\ref{fig:intro}a. We calculate the pressure eigenmodes and acoustic band structure of this system using the commercial finite-element software COMSOL Multiphysics{\textsuperscript{\textregistered}}.

\ptitle{Anticipation of harmonic modes}
Harmonic oscillator states form within a band dispersion that is locally parabolic in both real and momentum space (Fig.~\ref{fig:intro}b), mimicking the behavior described by equation (\ref{eq:ho}). The top band of the $s$-manifold of bilayer graphene has a kinetic energy dispersion $E(\mathbf{k}) \propto - |\mathbf{k}|^2$ around the $\Gamma$ point, as calculated in our honeycomb metamaterial (blue line in Fig.~\ref{fig:intro}c). Across the moir\'e supercell, the interlayer coupling $t_z$ is strongest when all the cavities perfectly overlap (AA stacking) and weakest when the overlap is smallest. This local variation in $t_z$ creates a position-dependent potential, $V(\mathbf{r})$, which can be approximated for small $\theta$ by the stacking configuration of two untwisted layers with a relative shift\cite{carrNRM2020}, as shown in Fig.~\ref{fig:intro}d. In our acoustic metamaterial, the $\Gamma$-point eigenfrequency of the upper band peaks at the AA, AB, and BA stacking configurations, corresponding to a local potential $V(\mathbf{r})\propto -|\mathbf{r}|^2$ (blue line in Fig.~\ref{fig:intro}d). We expect harmonic oscillator states to emerge within these local maxima because the $\mathbf{k}$-space curvature of the $\Gamma$-point band dispersion has the same sign as the $\mathbf{r}$-space moir\'e potential. This scenario leads to a local Hamiltonian equivalent to equation (\ref{eq:ho}) with a negative effective mass. The angular frequency of the harmonic oscillator can be approximated by \cite{CarrPRR2020}
\begin{equation}\label{eq:omega}
    \omega_0 = \sqrt{\dfrac{\partial^2 H(\mathbf{k},\mathbf{r})}{\partial \bf{k}^2}\dfrac{\partial^2 H(\mathbf{k},\mathbf{r})}{\partial \bf{r}^2}}.\\
\end{equation}
\noindent We estimate the $\mathbf{k}$ and $\mathbf{r}$ space curvatures by fitting parabolas to the blue band maxima in Fig.~\ref{fig:intro}c and d, yielding an expected level spacing of $\omega_{0} \approx 340$ Hz for the AA point from equation (\ref{eq:omega}).

\ptitle{HO states and their Q-factor}
By twisting the metamaterial as illustrated in Fig.~\ref{fig:intro}a, our simulation reveals a ladder of harmonic oscillator states localized around the AA sites, shown in Fig.~\ref{fig:aa_harmonic}a. These states form flat bands with roughly equal spacing $\omega_{0} \approx 340$ Hz, in excellent agreement with the independent approximation from the shifted metamaterial using equation (\ref{eq:omega}). The bands have extremely narrow intrinsic linewidths $\Delta f\sim 1 \ \mu$Hz, corresponding to a quality factor $\mathrm{Q} \equiv f/\Delta f$ exceeding $10^{10}$. However, when we account for realistic acoustic losses due to damping within air and steel, Q falls to around 4,000, which still exceeds typical measured Q for airborne acoustic devices at ambient conditions \cite{hashimotoSR2022,kronowetterNC2023}. By optimizing material selection, this design concept has the potential to achieve ultra-high Q.

\ptitle{Orbital analogy and degeneracy} 
In real space, the harmonic oscillator modes exhibit well-defined spatial symmetries, as shown in Fig.~\ref{fig:aa_harmonic}b. The spatial patterns are the angular harmonics of a two-dimensional quantum harmonic oscillator, and we label these modes as $s,$ $p,$ and $d$ orbitals, in analogy to the orbital states of electrons around a nucleus. In contrast, dispersive bands with non-localized real-space distributions form below the potential well, shown in Fig.~\ref{fig:aa_harmonic}a,c. As expected, the $s$ mode is non-degenerate, and the $p$ mode is two-fold degenerate. Meanwhile, the $d$ modes exhibit degeneracy breaking, with the $d_0$ mode splitting from the $d_{-1}$ and $d_{1}$ modes, likely due to the anharmonicity of the moir\'e potential well. Moreover, these harmonic oscillator states couple to create larger spatial arrays of high-Q modes, forming an emergent triangular lattice on the moir\'e length scale (Fig.~\ref{fig:aa_harmonic}d). 

\ptitle{AB-localized states}
Harmonic oscillator states also emerge at the AB and BA sites, where $V(\mathbf{r})$ has additional local maxima.  In this case, only the $s$ and $p$ AB/BA harmonic modes fit within the potential, highlighted as red lines in Fig.~\ref{fig:aa_harmonic}a. The level spacing $\omega_0 \approx 250$ Hz is smaller than the AA-localized modes, reflecting the wider local curvature of the potential around AB/BA (see equation (\ref{eq:omega})). In contrast to the AA-localized modes, which form a triangular lattice with nearest-neighbor distance $\sqrt{3} \lambda$, the composite AB/BA-localized modes form a honeycomb lattice with nearest-neighbor distance $\lambda$. 

\begin{figure}
\includegraphics[width=8.6cm]{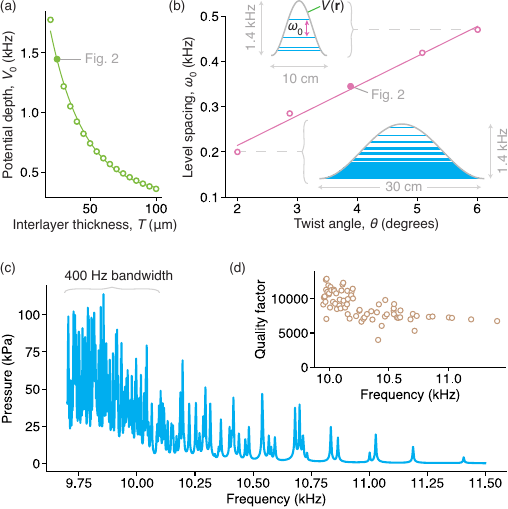}
\caption{\textbf{Tuning the properties of an acoustic harmonic resonator.} \textbf{a,} The depth, $V_0$, of the AA-centered potential well grows with reduced interlayer thickness. The thickness $T=25~\mu$m used in Fig.~\ref{fig:aa_harmonic} lies in the strong coupling regime where many harmonic oscillator states are visible. \textbf{b,} The spacing between the lowest two harmonic oscillator states, $\omega_0$, grows linearly with twist angle $\theta$, at fixed $T=25~\mu$m.  The potential (gray line in insets) widens at lower twist angles, allowing more harmonic oscillator modes (blue lines) to form. \textbf{c,} At $\theta = 2^{\circ},$ the ladder of harmonic oscillator states appears as a series of closely-spaced, sharp peaks in the pressure response amplitude as a function of source excitation frequency. The higher angular momentum modes become densely packed in frequency, creating an effective 400 Hz bandwidth resonance response. The absolute amplitude of the peaks is dependent on the method of excitation, in this case, eight point sources placed in cavities throughout the moir\'e cell. \textbf{d,} The corresponding Q values for the peaks in panel \textbf{c}, obtained by fitting a Lorentzian curve to each resonance, remain greater than 4000 for all modes.} 
\label{fig:tunability}
\end{figure}

\ptitle{Tunability of moir\'e potential depth} In the stacked two-dimensional semiconductor systems that inspired this work \cite{seyler2019, jin2019, peiPRX2022}, the moir\'e potential is determined by the microscopic interactions between electrons and atoms, limiting tunability. In contrast, the interlayer coupling in a metamaterial can be continuously tuned by adjusting the interlayer thickness \cite{dorrellPRB2020}. A stronger interlayer coupling increases the $\Gamma$-point band splitting, which deepens the harmonic oscillator potential well $V_0$, as shown in Fig.~\ref{fig:tunability}a. A deepened well allows for an extended ladder of harmonic oscillator modes, enabling access to those with higher angular momentum. 

\ptitle{Twist angle tunes potential width}
Our metamaterial also allows the width of the potential well to be tuned.  The potential $V(\mathbf{r})$ can be widened by decreasing the twist angle, which moves the AB points further away from the AA point, stretching the same change in $t_z$ over a longer length scale. A wider $V(\mathbf{r})$ reduces the curvature $\partial^2 H(\mathbf{r}) / \partial \mathbf{r}^2$, leading to more closely spaced resonances from equation (\ref{eq:omega}). The relationship between level spacing and twist angle is nearly linear, suggesting that harmonic oscillator modes become densely packed in frequency as the twist angle approaches zero (Fig.~\ref{fig:tunability}b). For instance, adjusting the twist angle from $6\degree$ to $2\degree$ increases the number of non-degenerate harmonic oscillator modes from 4 to 25 (inset in Fig.~\ref{fig:tunability}b). At a twist angle of $2\degree$, we found that although the $s$ and $p$ modes are separated by about 200 Hz, higher angular momentum modes can be spaced as little as 10 Hz apart. 
Crucially, this low-twist regime yields an acoustic resonator that maintains a high Q over a wide bandwidth (Fig.~\ref{fig:tunability}c,d), overcoming the typical trade-off between Q and bandwidth.

\ptitle{Conclusion} Acoustic twisted vdW metamaterials offer a promising platform for realizing broadband, high-Q resonance modes. Our twisted bilayer metamaterial reveals a ladder of harmonic oscillator modes localized around the AA, AB, and BA stacking configurations. The depth and width of the harmonic potential well depend on continuously tunable properties of the metamaterial, allowing the number and frequency separation of the resonances to be adjusted as desired. Unlike traditional acoustic fabrication techniques where binary potentials are encoded through sudden changes in material density, our approach inherits a smoother long-range potential from the moir\'e pattern. This smooth potential, combined with the strong localization and multi-cavity nature of these emergent modes, makes them resilient to local fabrication defects. We anticipate these resonators will advance applications requiring broadband, high-Q performance, such as efficient amplification for energy harvesting \cite{akbari-farahaniSaAAP2024}, chemical and biological sensing \cite{LuchanskyAnalChem2012}, and second-harmonic generation  \cite{KapfingerNatComm2015, TadesseNatComm2014, singhLSA2020, zhangNC2022}. Moreover, although our simulations focus on macroscopic length scales and kHz frequencies, the underlying principles of moir\'e localization extend naturally to surface acoustic\cite{yuNM2016,wangNC2022,achaouiPRB2011,kahlerCP2022} and photonic systems\cite{oudichPRB2021,bandresS2018, tangSA2023,liLSA2021}, opening doors for novel microwave and optical devices.


\vspace{7cm}
\newpage

\section{Methods}

\textbf{Metamaterial Geometry --} Our metamaterial contains two materials: steel plates (density 7850 kg/m$^3$, speed of sound 5790 m/s) and air cavities and channels (1.2 kg/m$^3$, 343 m/s). For all metamaterials, simulated in COMSOL Multiphysics{\textsuperscript{\textregistered}}, we use the physics-controlled ``fine'' mesh. Each monolayer metamaterial consists of a honeycomb lattice of air cavities, with cavity radius of $R = 3.25$ mm connected by air channels of width $w = 0.875$ mm with nearest-neighbor spacing $a = 10$ mm, surrounded by a $D = 1$ mm thick steel sheet. To form the bilayer metamaterial in Fig.~\ref{fig:aa_harmonic}, two such steel sheets are stacked and coupled by an interlayer membrane of thickness $T = 25$ $\mu$m, as well as two boundary steel sheets on the top and bottom to trap the acoustic waves. We choose a commensurate twist angle of $\theta = 3.89^{\circ}$, giving rise to a hexagonal moire supercell with three pairs of periodic Floquet boundary conditions on each side. The top and bottom sheets are impedance matched to air on their outer faces. \newline
\indent \textbf{Band Structure Calculations --}
We simulate our metamaterial models using the Pressure Acoustics, Frequency Domain interface within the acoustics module of COMSOL Multiphysics{\textsuperscript{\textregistered}}. We create an eigenfrequency study using a parametric sweep in $\bf{k}$-space around the supercell Brillouin zone along the path $M_s \to \Gamma_s \to K_s.$ To perform Q simulations, we run a Frequency Domain Study by placing a monopole point source in one of the center cavities with a free space reference power (RMS) of 0.01 W. Losses through air cavities are modeled under atmospheric attenuation, which is simulated at standard temperature and pressure conditions of 293 K and 1 atm. Losses in the steel are defined with a attenuation coefficient of $\alpha=$ 0.05 dB/m. We calculate Q by fitting a Lorentzian to the power spectral density and calculating the ratio of the amplitude to the full width at half maximum.

\section{Acknowledgements}

This work was supported by the National Science Foundation Grant No.\ OIA-1921199 and the Science Technology Center for Integrated Quantum Materials Grant No.\ DMR-1231319.

\end{document}